\documentclass[aps,prl,twocolumn,showpacs,groupedaddress]{revtex4-1}

\usepackage{graphicx}
\usepackage{dcolumn}
\usepackage{bm}
\usepackage{amssymb}
\begin{document}

\title{Higher-order Laguerre-Gauss mode generation and  interferometry for gravitational wave detectors}

\author{M. Granata}
\email{granata@apc.univ-paris7.fr}
\author{C. Buy}
\author{R. Ward}
\author{M. Barsuglia}
\affiliation{Laboratoire Astroparticule et Cosmologie (APC)\\ Universit\'e Paris Diderot - CNRS: IN2P3 - CEA: DSM/IRFU - Observatoire de Paris, 10 rue Alice Domon et L\'eonie Duquet, 75013 Paris, France}

\date{\today}

\begin{abstract}

We report on the first experimental demonstration of higher-order Laguerre-Gauss (LG$^\ell _p$) mode generation and interferometry using a method scalable to the requirements of gravitational wave (GW) detection. GW detectors which use higher-order LG$^\ell _p$ modes will be less susceptible to mirror thermal noise, which is expected to limit the sensitivity of all currently planned terrestrial detectors. We used a diffractive optic and a mode-cleaner cavity to convert a fundamental LG$^0_0$ Gaussian beam into an LG$^3_3$ mode with a purity of 98\%.  The ratio between the power of the LG$^0_0$ mode of our laser and the power of the LG$^3_3$ transmitted by the cavity was 36\%. By measuring the transmission of our setup using the LG$^0_0$, we inferred that the conversion efficiency specific to the LG$^3_3$ mode was 49\%. We illuminated a Michelson interferometer with the LG$^3_3$ beam and achieved a visibility of 97\%.

\end{abstract}

\maketitle

Gravitational waves (GWs) are ripples in the metric of spacetime that propagate at the speed of light and can act as carriers of astrophysical information. GWs emitted by nearby strongly gravitating systems (such as black hole or neutron star binaries) are expected to be detectable on Earth. The currently operating ground-based GW detectors such as Virgo \cite{VirgoStatus} and LIGO  \cite{LIGOstatus} are Michelson-based interferometers with Fabry-Perot cavities in the arms. These first generation detectors approximately reached their design sensitivities and completed several observational runs, but no detections have been reported so far. 

Planned upgrades to these GW detectors should significantly increase their sensitivity. The upgraded detectors (Advanced Virgo \cite{AdVirgo} and Advanced LIGO \cite{aLIGO}), as well as future GW detectors such as the Einstein Telescope \cite{3rdgenPunturo}, will be limited by mirror thermal noise in the central region of the detection band (around 10$^2$ Hz). This noise arises from fluctuations of mirror surfaces under the random motion of particles in coatings and substrates \cite{SaulsonTN, LevinTN}.

One option to decrease this noise is to resonate higher-order Laguerre-Gauss (LG$^\ell _p$) modes \cite{kogelnikLaser, siegmanBook} in the detector arm cavities, rather than the currently used LG$^0_0$ fundamental Gaussian mode. LG$^\ell _p$ modes are a complete set of solutions to the paraxial wave equation, and their complex amplitude is given by \cite{kogelnikLaser, siegmanBook}
\begin{equation}
	\label{LGpl}
		\begin{array}{ll}
			u ^\ell _p(r,\phi,z) =  \sqrt{\frac{2P}{\pi}}\sqrt{\frac{p!}{(p+|\ell|)!}} \  \frac{1}{w(z)} \  \exp\Bigl[{\frac{-r^2}{w^2(z)}}\Bigr] \\
 			 \\ 
 			 \;\;\;\;\; \times  \Bigl(\frac{2r^2}{w^2(z)}\Bigr)^{|\ell|/2} \  L ^{|\ell|} _p\Bigl(\frac{2r^2}{w^2(z)}\Bigr) \  \exp\Bigl[-i\ell\phi\Bigr] \\
 			 \\ 
 			 \;\;\;\;\; \times \ \exp\Bigl[-i\Bigl(k\Bigl(z+\frac{r^2}{2R(z)}\Bigr)-(2p+|\ell|+1)\Phi _G\Bigr)\Bigr],
		\end{array}
\end{equation}
where $p$ and $\ell$ are the radial and the azimuthal indices respectively, $w(z)$ is the beam radius and $R(z)$ is the phase front curvature, $\Phi _G = \arctan(z\lambda/\pi w^2_0)$ is the Gouy phase, $w_0 = w(z=0)$ is the beam waist, and $L ^\ell _p(x)$ is the Laguerre generalized polynomial. LG$^\ell _p$ beams with $\ell \neq 0$ have $p+1$ radial nodes and spiral phase fronts, carrying orbital angular momentum of $\ell\hbar$ per photon \cite{allenMomentum}.

For the same mirror diameter and for equivalent diffraction losses and integrated power $P$, higher-order LG$^\ell _p$ beams have a multi-ringed power distribution which is wider than the distribution of the LG$^0_0$ mode. Because of this wider intensity distribution, higher-order LG$^\ell _p$ beams can average the thermal noise fluctuations over a bigger portion of the mirror surface, thus decreasing the impact of thermal noise \cite{calcMours}. Mesa (flat-top) beams have also been investigated \cite{DambroMesa} and generated in a prototype cavity \cite{TaraMesa}. Nevertheless, although mesa beams and other beam profiles may provide more noise reduction \cite{coneBondarescu, calcVinet}, higher-order LG$^\ell _p$ modes are attractive because they resonate in a cavity composed of spherical mirrors, a well established technology.   

It has been analytically demonstrated \cite{calcVinet, calcMours} that the thermal noise of the mirrors can be reduced by a factor which depends on the spatial order $N=2p+\ell$ of the LG$^\ell _p$ mode resonant in the interferometer: higher $N$ values lead to larger beams and lower thermal noise. For example, the thermal noise level could be decreased by nearly a factor of 2 by using an LG$^3_3$ beam \cite{calcVinet}. If the detector sensitivity is largely dominated by thermal noise, such an improvement can allow an increase of the detection horizon by approximately the same factor and hence the accessible volume of the Universe by its cube. 
\begin{figure}
	\centering
	\includegraphics[width=0.45\textwidth]{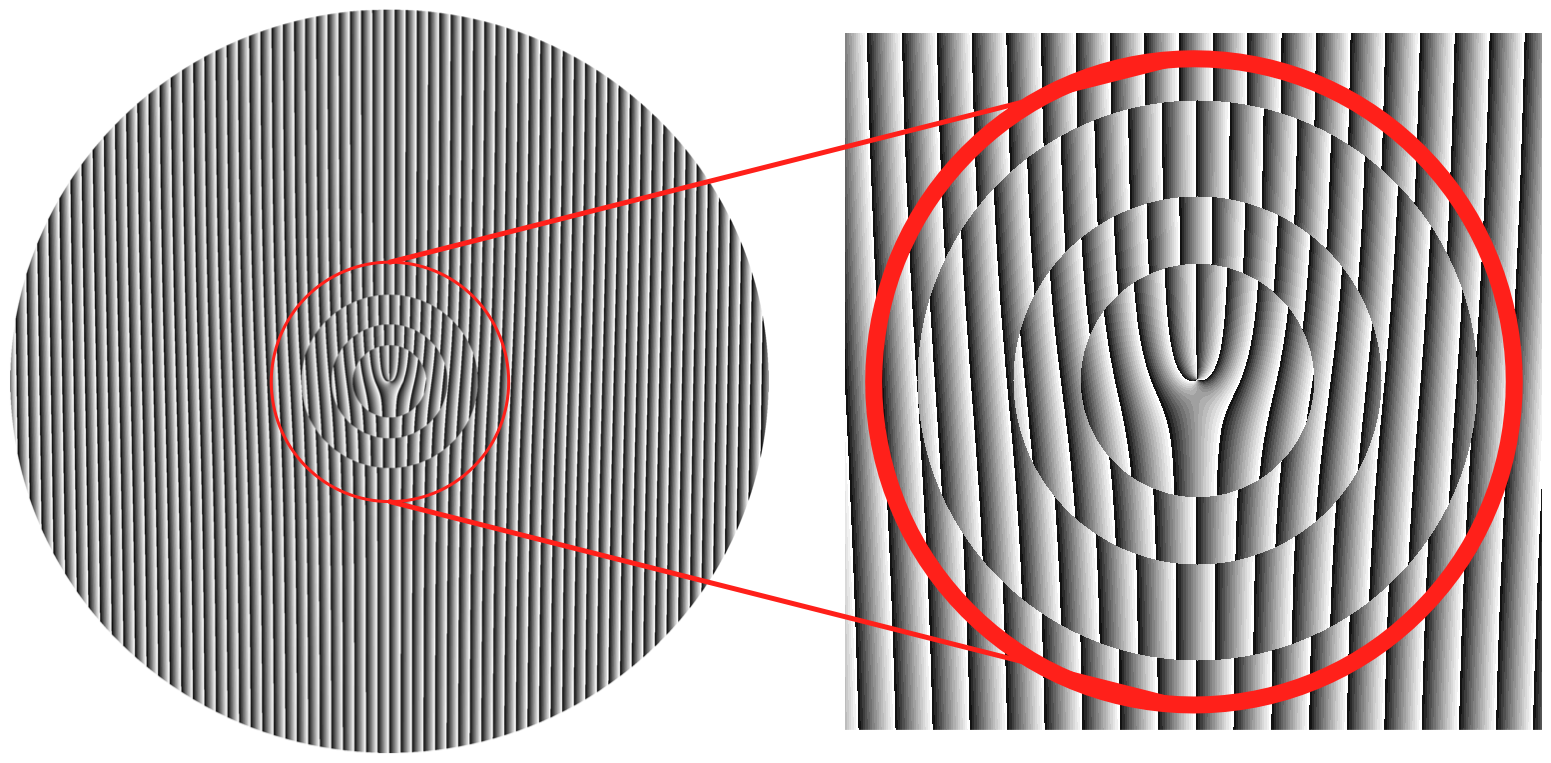} 
	\caption{\label{DOEpattern} (Color online) DOE pattern for the generation of an LG$^3_3$ mode.}
\end{figure}
Furthermore, present interferometric GW detectors are strongly affected by thermal issues, caused by laser power absorption in the optics (either in the bulk or on the coating): the absorbed power gives rise to a temperature gradient in the material, which results in a refractive index change and in a thermal deformation of the mirror surface. The resulting aberrations in the beam wavefront cause a loss of detector sensitivity. Thermal effects related to higher-order LG$^\ell _p$ beams should be in general lower than those given by the Gaussian intensity pattern of the LG$^0_0$ \cite{calcVinet}.

To be used in GW detectors, higher-order LG$^\ell _p$ beams must be generated with very high purity and stability. The mode purity is crucial for having far-field propagation in kilometer-scale interferometers with no degradation of the propagating beam shape, and for optimal coupling of the mode to the Fabry-Perot cavities of the detector. Moreover, since high-power laser beams of hundreds of Watts will be used, higher-order LG$^\ell _p$ modes must be generated with high efficiency and low losses.

Although higher-order LG$^\ell _p$ modes have been used for quantum optics experiments \cite{entanglMair} and optical tweezers \cite{tweezGrier}, to our knowledge the only application proposed in high-precision optical interferometry is GW detection. A tabletop experiment demonstration of this technique is then needed before its implementation on kilometric scale detectors. Preliminary investigations of LG$^\ell _p$ interferometry have been reported in \cite{LG33interf}. In this Letter we report on the first demonstration of high-purity LG$^3_3$ mode generation and interferometry which uses a technique compatible with future GW interferometers.

Many techniques are presently available to generate a higher-order LG$^\ell _p$ mode from a laser emitting an LG$^0_0$ beam. Cylindrical mode converters \cite{CMCs} can transform higher-order Hermite-Gauss beams \cite{siegmanBook} into higher-order LG$^\ell _p$ beams, using astigmatic lenses; spiral phase plates \cite{SPPs} are optics whose varying thickness induce the typical LG$^\ell _p$ spiralling phase pattern into the input beam; diffractive optics, which include computer generated holograms \cite{CGHs}, spatial light modulators \cite{SLMs} and etched-glass diffractive optics \cite{DOEs} can modulate the amplitude and the phase of the incoming beam to obtain a higher-order LG$^\ell _p$ mode.

Glass-made diffractive optical elements (DOEs) in particular seem to be the simplest and most suitable solution for interferometric GW detectors. A DOE is an etched waveplate that acts as general wavefront transformer, allowing the direct conversion of a fundamental LG$^0_0$ mode into a higher-order LG$^\ell _p$ beam (for a more detailed description of the working principle see for example \cite{DOEs}). Moreover, glass-made DOEs are stable passive optics which can handle the high power that will be necessary for future detectors.

We used a commercial 1064 nm Nd:YAG NPRO laser and a fused silica DOE designed by SILIOS Technologies \cite{SILIOSsite} for the generation of an LG$^3_3$ beam. This DOE has 2400$\times$2400 pixels, each one measuring 5.9 $\mu$m, and 16 levels of phase are etched on its surface. Its phase pattern is obtained by starting from the theoretical phase of an LG$^3_3$ mode, and then optimizing this phase to improve the shape of the generated mode. A blazed grating pattern is superimposed to this phase pattern to remove the unmodulated components of the output beam and increase the generated beam quality. The etched pattern of this DOE, measuring about 15 mm in diameter, is shown on Fig.\ref{DOEpattern}. The radius of the LG$^0_0$ beam impinging on the DOE is 2.15 mm. The measured transmitted power on the diffraction order of interest is more than 80\%. The application of an anti-reflective coating should increase this value.
\begin{figure}
	\centering
	\includegraphics[width=0.50\textwidth, height=0.16\textheight]{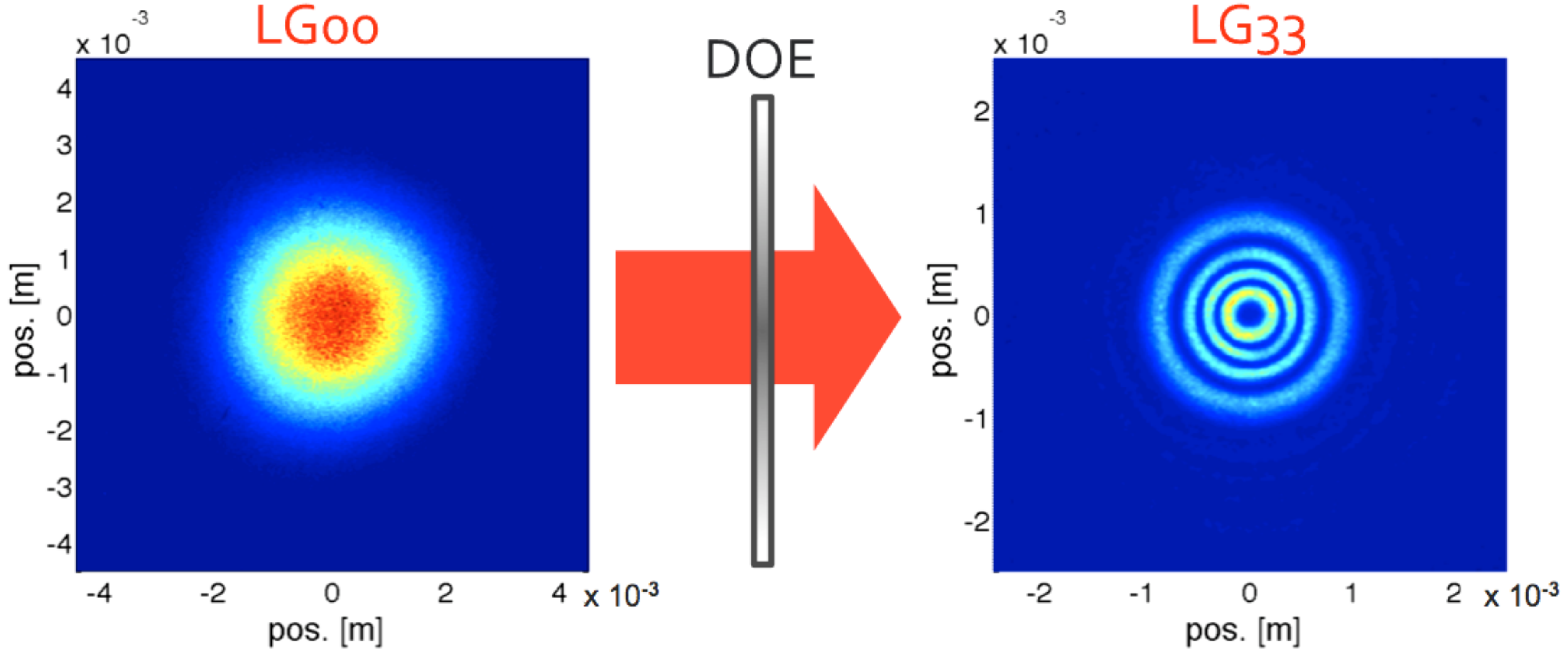} 
	\caption{\label{genMode} (Color online) DOE working principle: the input LG$^0_0$ beam (left) is converted into an LG$^3_3$ beam (right); the distributions shown here are acquired before and after the DOE. The plots have the same color scale.}
\end{figure}

We estimate the purity of our DOE-generated LG$^3 _3$ mode by using a 2-dimensional amplitude overlap integral $\gamma$, calculated as the scalar product between the measured and the theoretical distributions, indicated as $LG^3_3|_{measure}$ and $LG^3_3|_{theory}$ respectively:
\begin{equation}
	\label{OvInt}
	\gamma = \big\langle LG^3_3|_{theory} \big|  LG^3_3|_{measure} \big\rangle,
\end{equation}
where $LG^3_3|_{theory}$ is given by the modulus of Eq. (\ref{LGpl}) for a given beam size $w(z)$ and a given distance $z_0$ from waist. Since in Eq. (\ref{OvInt}) the phase of the two modes is neglected, this definition should be considered as an upper-limit estimation of the purity of $LG^3_3|_{measure}$. For our DOE-generated LG$^3_3$ beam, shown in Fig.\ref{genMode}, we have $\gamma = 88\%$. This value of purity can be partly explained by the intrinsic conversion efficiency of the DOE, which can be further improved by optimizing the DOE design, and also by our slightly astigmatic LG$^0_0$ input beam. From the overlap integral (\ref{OvInt}) we can compute the {\it coupling losses} $L$, i. e. the power which is not converted into the LG$^3_3$ mode, defined as $L=1-\gamma^2 = 23\%$.

\begin{figure}
	\centering
	\includegraphics[width=0.50\textwidth, height=0.25\textheight]{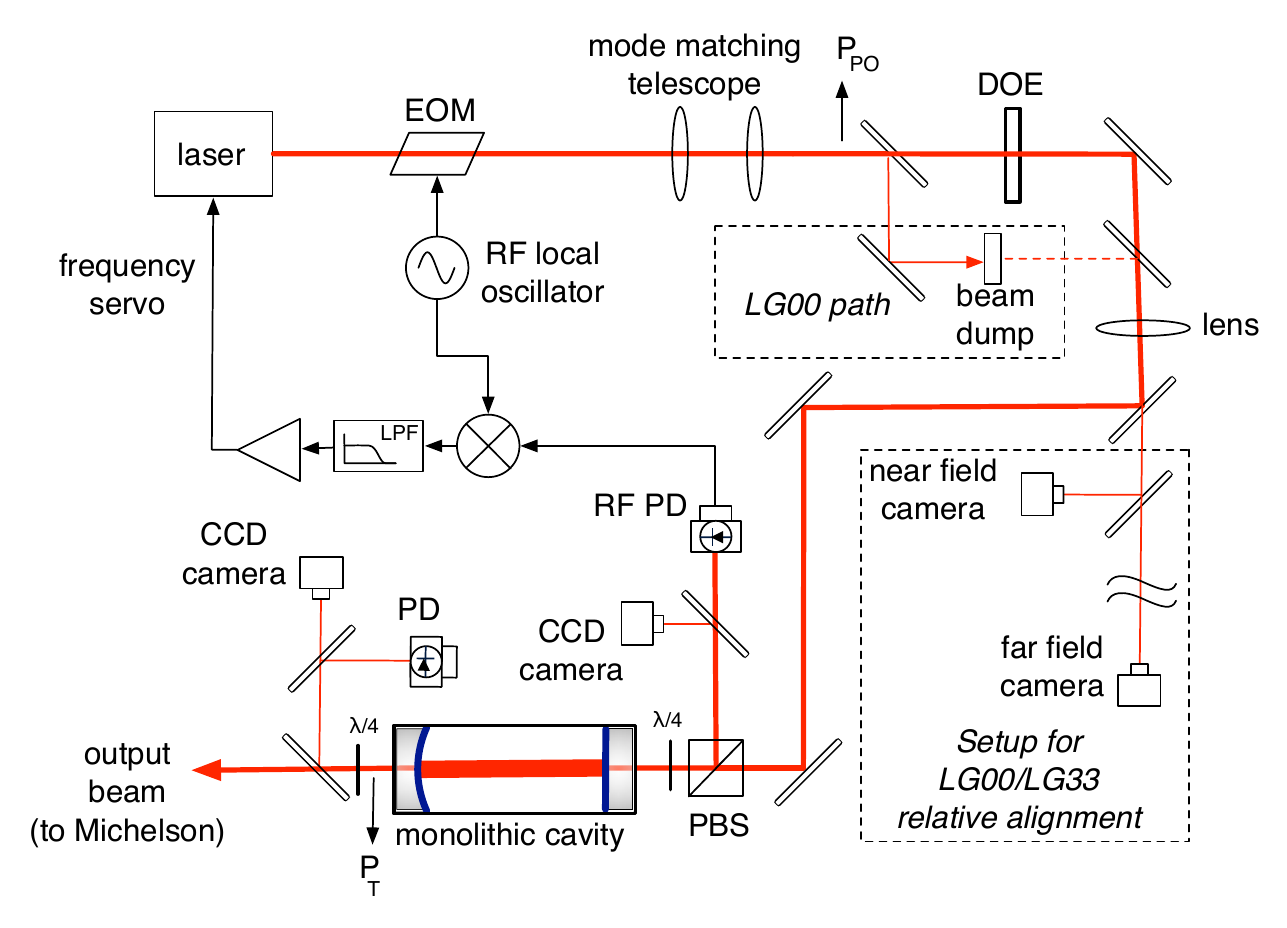} 
	\caption{\label{expSetup} (Color online) Experimental setup used for the generation of the LG$^3_3$ mode.}
\end{figure}

In order to increase the generated mode purity, DOE-generated modes must be spatially filtered: this is achievable by using a mode-cleaning filter cavity \cite{FPMC}. To test the performance of this spatial filtering technique, we assembled a tabletop setup for the injection of the generated LG$^3_3$ mode into a linear mode-cleaner cavity. A schematic of the experimental setup is shown in Fig. \ref{expSetup}. The LG$^0_0$ beam from the laser is passed through an electro-optic modulator for the generation of radio-frequency sidebands and then goes through the DOE.  The generated LG$^3_3$ mode is then sent to the mode-cleaner, which is a 30 cm long plano/concave monolithic cavity with a finesse of 100.  At the LG$^3_3$ resonance, the frequency of the laser is locked to the mode-cleaner length using a standard Pound Drever-Hall locking scheme \cite{PDHlock}. 

All the optics of the setup are first aligned using the fundamental LG$^0_0$ mode of the laser, which can propagate on a pick-off path without going through the DOE. In our setup either the LG$^0_0$ or the LG$^3_3$ beam can be blocked independently. This allows the alignment of the LG$^3_3$ beam on the monolithic cavity with the following procedure: (i) the LG$^3_3$ beam is blocked and the LG$^0_0$ is aligned to the mode-cleaner; (ii) the LG$^3_3$ is superimposed to the LG$^0_0$ using two CCD cameras placed at different locations, in near-field and in far-field regime; (iii) the LG$^0_0$ beam is blocked and the alignment of the LG$^3_3$ is tuned using the beam reflected from the cavity, monitored by a dedicated CCD camera.

We have been able to lock the laser on the LG$^3_3$ resonance for several hours, and for the whole measurement time the DOE has shown a very stable behaviour. We could monitor the beam transmitted by the mode-cleaner using a CCD camera and a photodiode. At the LG$^3_3$ resonance, we measured the beam power $P_{in}$ going into the mode-cleaner and the transmitted power $P_{out}$ coming out from the cavity: the resulting transmission is $P_{out}/P_{in}= 58 \%$. The cavity throughput $\tau = 90\%$ has been measured separately by injecting the LG$^0_0$ mode.

\begin{figure}
	\centering
	\includegraphics[width=0.42\textwidth, height=0.17\textheight]{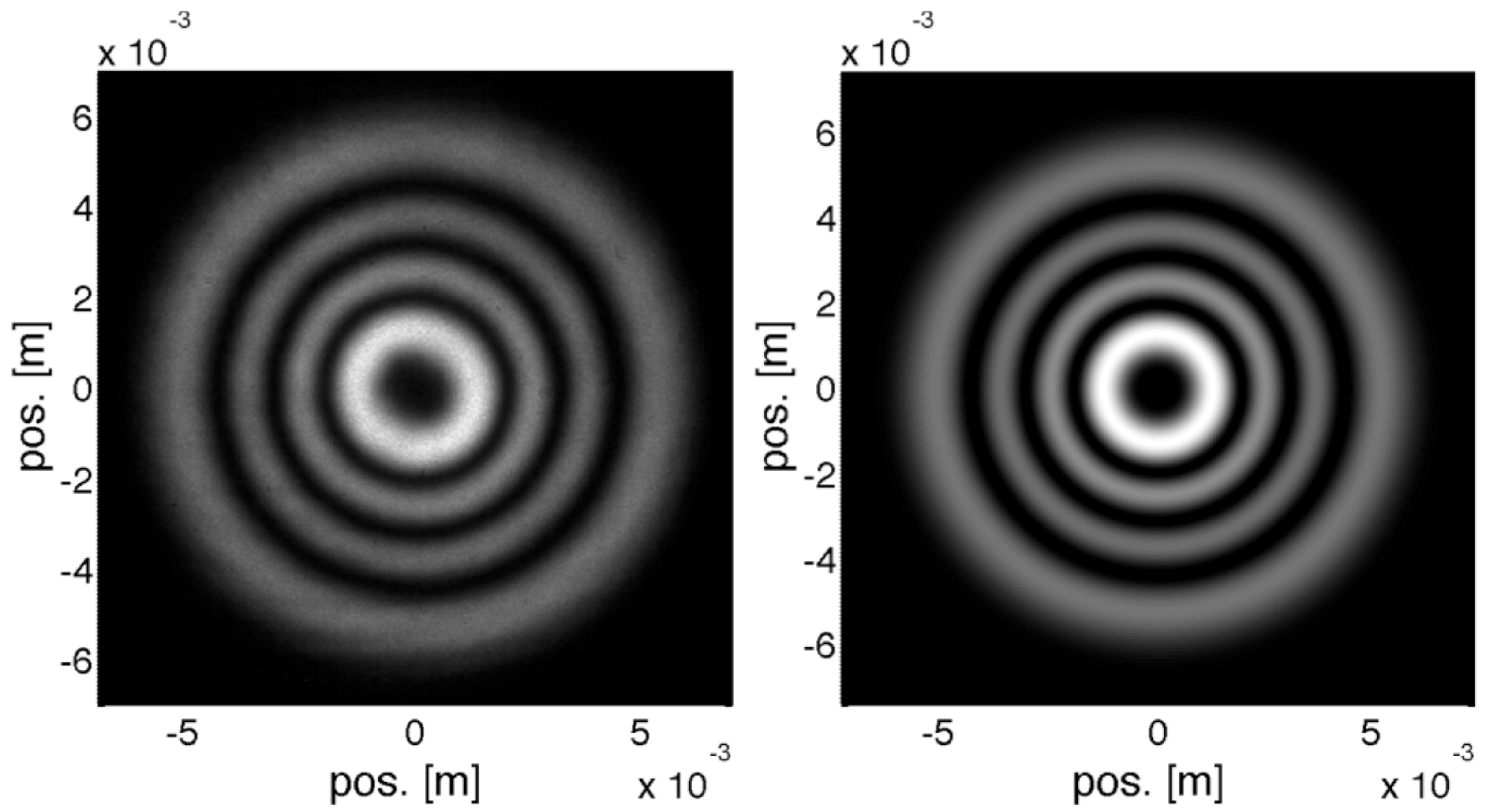} 
	\caption{\label{modeComp} Transverse intensity distribution comparison: measured LG$^3_3$ distribution (left) at $z_0=1.30$ m from beam waist; expected LG$^3_3$ theoretical distribution for $z_0/z_R=5.4$ (right). The plots have the same color scale. }
\end{figure}

The filtered LG$^3_3$ mode transmitted by the mode-cleaner is shown in the left panel of Fig.\ref{modeComp}: it has been measured in far-field regime, when the distance $z_0$ from the beam waist is much bigger than the Rayleigh range $z_R=\pi w^2_0/\lambda$. In our case, $z_0=$ 1.30 m and $z_0/z_R=$ 5.4. This mode is compared with the corresponding expected analytical intensity distribution, calculated by propagating the ideal cavity mode over $z_0$, which yields the field shown in the right panel of Fig.\ref{modeComp}. The purity computed with the overlap integral for the two distributions of Fig.\ref{modeComp} has increased up to $\gamma = 98 \%$, and the coupling losses have decreased to $L = 4\%$. 
 
\begin{figure}
	\centering
	\includegraphics[width=0.45\textwidth, height=0.19\textheight]{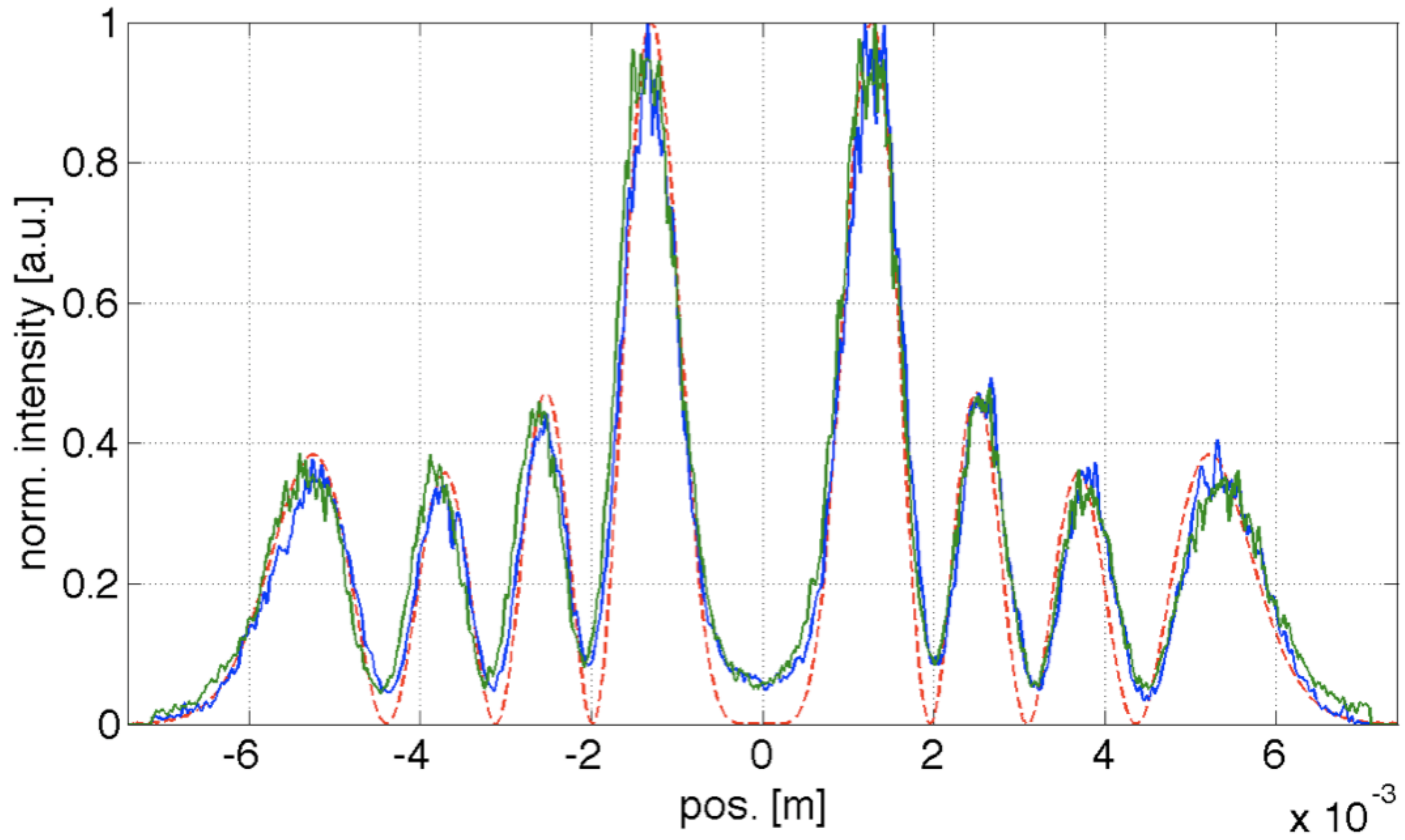} 
	\caption{\label{profComp} (Color online) Cross section comparison: expected cross section for $z_0/z_R=5.4$ (red dashed line), measured horizontal (blue solid line) and vertical (green solid line) cross sections at $z_0=1.30$ m from beam waist.}
\end{figure}

The cross sections of the intensity patterns shown in Fig.\ref{modeComp} are compared in Fig.\ref{profComp} (where the speckle effect on the measured profiles is due to the camera we used for the acquisition). The main difference between the cross sections is the presence of some power left at the radial nodes, but these details will likely be improved by increasing the finesse of the mode-cleaner cavity. Apart from that, the observed pattern presents an overall correspondence with the expected profile, and this also indicates that the divergence of the filtered LG$^3_3$ beam is close to that of the theoretical cavity mode.

We estimated the LG$^0_0$ to LG$^3_3$ conversion efficiency as follows. We first measured the ratio  $(P_{33})_T / P_{PO}$= 36\%, where $P_{PO}=435$ mW is the power of the LG$^0_0$ beam at the input of the LG$^0_0$ pick-off (as indicated by the arrow on Fig. \ref{expSetup}) and $(P_{33})_T = 155$ mW is the power of the LG$^3_3$ beam transmitted by the mode-cleaner at the resonance. Then, using the LG$^0_0$ beam, we measured the transmission $\eta = 78\%$ of our setup, where losses are caused by the power fraction that goes into the LG$^0_0$ pickoff (not used in the LG$^3_3$ generation) and by the optical loss of all the optics (mirrors, lens, polarizer cube) placed before the mode-cleaner. From these measurements we infer that the conversion efficiency is $\epsilon = \gamma^2 (P_{33})_T/(P_{00})_T = 49\%$, where $(P_{00})_T = \tau\eta P_{PO}$ is the LG$^0_0$ power transmitted by our setup through the mode-cleaner and $\gamma^2 = 96\%$ is the transmitted LG$^3_3$ mode content. Of the 51\% power lost in the conversion, about 20\% is lost at the DOE, going in scattered light (due to the absence of the anti-reflective coating) and in higher diffraction orders. The rest of the power is reflected by the cavity when it is locked on the LG$^3_3$ mode, in the form of modes of different order. A further increase of efficiency could be achieved by optimizing the DOE pattern and by tuning the matching between the beam and the mode-cleaner cavity. The transmission $\eta$ could be largely improved by replacing the LG$^0_0$ pick-off with a movable flip-mount, and by using high quality, low loss optics in our setup.

The filtered LG$^3_3$ beam is injected into a Michelson interferometer, whose arms are 15 cm long. A piezoelectric actuator, glued on one of the two arm mirrors, provides the actuation for locking the interferometer on the dark fringe working point. At the anti-symmetric port of the interferometer, a photodiode detects the interference signal coming from the beamsplitter, and this signal is used in a servo for the locking control. By manually tuning optics alignment we obtained a fringe visibility of 97\%.

In summary, we developed a technique to generate a high-purity LG$^3 _3$ beam which should be scalable to the power required in future interferometric GW detectors. The generated LG$^3 _3$ mode has a purity of 98\%, and the power of the LG$^3_3$ mode transmitted by the mode-cleaner is 36\% of that delivered by our laser on the LG$^0_0$ mode. We infer that the conversion efficiency specific to the LG$^3_3$ mode is 49\%. Any quantum noise increase due to the reduction of available power could be compensated by either a more powerful laser and/or the injection of squeezed light \cite{squeezGoda} into the output port. However, further improvements of the setup are ongoing in order to increase both the conversion efficiency and the purity of the filtered LG$^3 _3$ mode. 

The aligment and lock of the mode-cleaner cavity and of the Michelson interferometer demonstrate experimentally the feasibility of higher-order LG$^\ell _p$ mode interferometry, which is of particular interest for GW detection. Our interferometer will be upgraded soon to a power-recycled Michelson with Fabry-Perot arm cavities, aiming to test more quantitatively the main issues of a full GW detector configuration. These include the validation of a sensing scheme for longitudinal and angular control and the characterization of the sensitivity of the LG$^3 _3$ mode to mirror misalignments. Moreover, the influence of degeneracy of LG$^\ell _p$ modes of the same order $N=2p+\ell$ \cite{calcMours} (not present for the LG$^0_0$ mode) on the optical performances of the interferometer has to be assessed.

This research work has been performed with the support of the CNRS programme Particules et Univers and of the programme Physique des 2 Infinis (P2I). R.W. is supported by the European Gravitational Observatory. The authors kindly acknowledge the LISA team at APC for its contribution to this work, and E. Chassande-Mottin, M. Evans and L. Barsotti for their comments on the manuscript.


\begin{thebibliography}{100}

\bibitem{VirgoStatus} F. Acernese {\it et al.} (Virgo Collaboration), Class. Quantum Grav. {\bf 25}, 184001 (2008)
\bibitem{LIGOstatus} B. P. Abbott {\it et al.} (LIGO Scientific collaboration), Rep. Prog. Phys. {\bf 72}, 076901 (2009)
\bibitem{AdVirgo} F. Acernese {\it et al.} (Virgo Collaboration), Virgo internal note VIR-0027A-09, https://pub3.ego-gw.it/itf/tds/
\bibitem{aLIGO} J. R. Smith (for the LIGO Scientific collaboration), Class. Quantum Grav. {\bf 26}, 114013 (2009)
\bibitem{3rdgenPunturo} M. Punturo {\it et al.}, Class. Quantum Grav. {\bf 27}, 194002 (2010)
\bibitem{SaulsonTN} P. R. Saulson, Phys. Rev. D {\bf42}, 2437 (1990)
\bibitem{LevinTN} Yu. Levin, Phys. Rev. D {\bf 57}, 659 (1998)
\bibitem{kogelnikLaser} H. Kogelnik and T. Li, Appl. Opt. {\bf 5}, 1550 (1966)
\bibitem{siegmanBook} A. E. Siegman {\it Lasers} (University Science Books, Sausalito, California, 1986)
\bibitem{allenMomentum} L. Allen {\it et al.}, Phys. Rev. A {\bf45}, 8185 (1992)
\bibitem{calcMours} B. Mours {\it et al.}, Class. Quantum Grav. {\bf 23}, 5777 (2006)
\bibitem{DambroMesa} E. D'Ambrosio, Phys. Rev. D {\bf 67}, 102004 (2003)
\bibitem{TaraMesa} M. Tarallo {\it et al.}, Appl. Opt. {\bf 46}, 6648 (2007)
\bibitem{coneBondarescu} M. Bondarescu {\it et al.}, Phys. Rev. D {\bf 78}, 082002 (2008)
\bibitem{calcVinet} J. Y. Vinet, Living Rev. Relativity {\bf 12}, 5 (2009)
\bibitem{entanglMair} A. Mair {\it et al.}, Nature {\bf 412}, 313 (2001)
\bibitem{tweezGrier} D. G. Grier, Nature {\bf 424}, 810 (2003)
\bibitem{LG33interf} P. Fulda {\it et al.}, Phys. Rev. D {\bf 82}, 012002 (2010)
\bibitem{CMCs} M. W. Beijersbergen {\it et al.}, Opt. Comm {\bf 96}, 123 (1993)
\bibitem{SPPs} G. A. Turnbull {\it et al.}, Opt. Comm. {\bf 127}, 183 (1996)
\bibitem{CGHs} J. Arlt {\it et al.}, J. Mod. Opt. {\bf 45}, 1231 (1998)
\bibitem{SLMs} N. Matsumoto {\it et al.}, J. Opt. Soc. Am. A {\bf25}, 1642 (2008) 
\bibitem{DOEs} S. A. Kennedy {\it et al.}, Phys. Rev. A {\bf 66}, 043801 (2002)
\bibitem{SILIOSsite} SILIOS Technologies website: http://www.silios.com/
\bibitem{FPMC} B. Willke {\it et al.}, Opt. Lett. {\bf 23}, 1704 (1998)
\bibitem{PDHlock} R. W. P. Drever {\it et al.}, Appl. Phys. B: Photophys. Laser Chem. {\bf 31}, 97 (1983)
\bibitem{squeezGoda} K. Goda {\it et al.}, Nat. Phys. {\bf 4}, 472 (2008) 

\end{thebibliography}
\end{document}